\documentclass[aps,pra,twocolumn]{revtex4}
\usepackage{amsfonts}
\usepackage{amsmath}
\usepackage{amssymb}
\usepackage{graphicx}
\usepackage[english]{babel}
\usepackage{hyperref}
\usepackage[usenames,dvipsnames]{color}
\usepackage{babel}

\begin{document}

\title{Dynamical Casimir effect via four- and five-photon transitions using
strongly detuned atom}
\author{A. V. Dodonov }
\email{adodonov@fis.unb.br}
\affiliation{Institute of Physics and International Center for Condensed Matter Physics,
University of Brasilia, 70910-900, Brasilia, Federal District, Brazil}

\begin{abstract}
The scenario of a single-mode cavity with harmonically modulated frequency
is revisited in the presence of strongly detuned qubit or cyclic qutrit. It
is found that when the qubit frequency is close to $3\nu $ there is a peak
in the photon generation rate via four-photon transitions for the
modulation frequency $4\nu $, where $\nu $ is the average cavity frequency.
Effective five-photon processes can occur for the modulation frequency $5\nu $ in the presence of a cyclic
qutrit, and the corresponding transition rates exhibit series of peaks. Closed analytical description is derived for the unitary evolution, and numeric simulations indicate the feasibility of multi-photon dynamical Casimir effect under weak dissipation.
\end{abstract}

\maketitle

\section{Introduction}

\label{s1}

The problem of photon generation from vacuum in response to fast variations
of the geometry or material properties of some resonator has been
extensively studied since the decade of 1970 \cite{moore} and became known
as the dynamical Casimir effect (DCE) (see the reviews \cite%
{rev1,rev2,rev3,rev4} for details). The main role of the resonator is to
enhance the photon creation \cite{lambre1,lambre2}, as DCE also takes place
in free space due to nonuniform acceleration of mirrors or dielectric bodies
\cite{d1,d2,d3}. In 2012 DCE was implemented experimentally in a microwave
cavity using a Josephson metamaterial, where the cavity effective length was
modulated by external magnetic flux \cite{meta}.

The mechanism responsible for the photon generation can be understood from
the paradigm of a single-mode cavity with an externally prescribed
time-dependent frequency $\omega (t)$. As shown in Ref. \cite{law1}, within
the framework of instantaneous mode functions and the associated dynamical
Fock space, the dynamics of the cavity field can be described by the
effective Hamiltonian $\hat{H}/\hbar =\omega (t)\hat{a}^{\dagger }\hat{a}%
+i\chi (t)(\hat{a}^{\dagger 2}-\hat{a}^{2})$, where $\hat{a}$ and $\hat{a}%
^{\dagger }$ are the instantaneous annihilation and creation operators and
(in the simplest case) $\chi =(4\omega )^{-1}d\omega /dt.$ The resulting
dynamics resembles the well known phenomenon of parametric amplification
\cite{rev4}, namely, photon pairs are generated from vacuum for the harmonic
perturbation $\omega (t)=\nu +\varepsilon \sin (\eta t)$, where $\nu $ is
the unperturbed cavity frequency, $\varepsilon $ is the amplitude and $\eta
=2\nu $ is the frequency of modulation \cite{rev1}. Photon pairs can also be
generated for fractional frequencies $2\nu /k$ due to higher harmonics (for
nonmonochromatic modulation \cite{do}) or $k$-th order effects (for harmonic
perturbation \cite{tom}), where $k$ is a positive integer. Moreover, when
the cavity field is coherently coupled to other quantum subsystems (e. g.,
multi-levels atom or harmonic oscillators \cite{pa1,pa2,pa3}) photons can be
generated (or annihilated \cite{diego,lucas}) for several other modulation
frequencies at the cost of entangling the subsystems. In particular, it has been recently predicted that a
dispersive cyclic qutrit \cite{cQED7,cycl1,cycl2,cycl3,cycl4} with
time-dependent energy splittings permits the generation of photons from
vacuum for $\eta \approx \nu $ and $\eta \approx 3\nu $ via effective one-
and three-photons transitions, respectively \cite{hara}.

In this paper it is shown that photons can also be generated from vacuum for
$\eta\approx 4\nu$ and $\eta\approx 5\nu$ (via effective four- and
five-photons transitions) by placing into the oscillating cavity a strongly
detuned qubit and cyclic qutrit, respectively. For brevity, these phenomena
are called 4- and 5-photon dynamical Casimir effects (4DCE and 5DCE), since
the stationary atom remains approximately in the ground state during the
evolution. The overall behavior does not depend on the precise dependence of
$\chi$ on $\omega$, so for simplicity it is assumed $\chi
=(4\omega)^{-1}d\omega /dt$ throughout the paper. The photon creation rates
are usually very small, however, it is predicted analytically and confirmed
numerically that they increase orders of magnitude in the vicinity of
certain atomic frequencies, becoming of the order $10^{-3}\varepsilon$. The
analytical description of the unitary dynamics is derived in the
dressed-states basis, and it is shown that for a constant modulation
frequency the amount of created photons is limited due to effective Kerr
nonlinearities.

This paper is organized as follows. General analytical description of the
unitary dynamics is presented in Sec. \ref{s2}. In Sec. \ref{s3} the case of
a qubit is studied in detail, and approximate expressions for the 4DCE
transition rate are derived in different regimes of parameters. The dressed-states of the cyclic qutrit and the
resonant enhancement of 5DCE are discussed in Sec. \ref{s4}. Sec. %
\ref{s5} presents exact numeric results on the system dynamics for the
initial vacuum state, confirming the analytical predictions and exemplifying
typical behaviors of 4DCE and 5DCE; the influence of dissipation is also
briefly discussed for the case of a qubit in Sec. \ref{s5a}. Finally, the
conclusions are summarized in Sec. \ref{s6}.

\section{General description}

\label{s2}

Consider a single cavity mode with time-dependent frequency $\omega (t)=\nu
+\varepsilon \sin (\eta t)$ that interacts with a qutrit in the cyclic
configuration \cite{cQED7,cycl1,cycl2,cycl3,cycl4}, so that all the atomic
transitions are allowed via one-photon transitions. The Hamiltonian reads%
\begin{eqnarray}
\hat{H}/\hbar &=&\omega \hat{n}+i\chi \left( \hat{a}^{\dagger 2}-\hat{a}%
^{2}\right) +\sum_{k=1}^{2}E_{k}\hat{\sigma}_{k,k}  \notag \\
&&+\sum_{k=0}^{1}\sum_{l>k}^{2}G_{k,l}(\hat{a}+\hat{a}^{\dagger })(\hat{%
\sigma}_{l,k}+\hat{\sigma}_{k,l})\ ,  \label{h0}
\end{eqnarray}%
where $\hat{a}$ ($\hat{a}^{\dagger }$) is the cavity annihilation (creation)
operator and $\hat{n}=\hat{a}^{\dagger }\hat{a}$ is the photon number
operator \cite{asu}. The atomic levels are $E_{0}\equiv 0,E_{1}$ and $E_{2}$%
, the corresponding states are denoted as $|\mathbf{k}\rangle $ and $\hat{%
\sigma}_{k,j}\equiv |\mathbf{k}\rangle \langle \mathbf{j}|$; to shorten the
final expressions the dipole-interaction term is abbreviated as $\hat{D}%
_{k,l}\equiv (\hat{a}+\hat{a}^{\dagger })(\hat{\sigma}_{l,k}+\hat{\sigma}%
_{k,l})$. The parameters $G_{k,l}\ll \omega $ denote the coupling strengths
between the atomic states $|\mathbf{k}\rangle $ and $|\mathbf{l}\rangle $
mediated by the cavity field, and it is employed a shorthand notation $%
G_{1}\equiv G_{0,1}$, $G_{2}\equiv G_{1,2}$ and $G_{3}=G_{0,2}$. Notice
that the {\em counter-rotating terms} are included in $\hat{H}$, otherwise the
effects presented in this paper disappear.

For weak modulation, $\varepsilon \ll \nu $, to the first order in $%
\varepsilon $ one has $\chi \approx (4\nu )^{-1}\varepsilon \eta \cos \eta t$%
. For the sake of generality the coupling strengths are also allowed to vary
with time as
\begin{equation}
G_{i}=g_{i}+\tilde{\varepsilon}_{i}\sin \left( \eta t+\phi
_{i}\right)\,,\quad i=1,2,3 \,,  \label{s}
\end{equation}%
where the phases $\phi_i$ are arbitrary. Such time-dependence may arise from
the primary mechanism of atom-field interaction, or be input externally \cite%
{libe}. For example, in the case of a stationary qubit (when only $G_{1}\neq
0$), the standard quantization in the Coulomb gauge \cite{schleich} gives $%
G_{1}\propto \sqrt{\omega } $, so to the first order in $\varepsilon $ one
gets $\phi _{1}=0$ and $\tilde{\varepsilon}_{1}=g_{1}\varepsilon /2\nu $.
This relationship will be used in Sec. \ref{s3} to illustrate the influence
of eventual variation of $G_i$, although the precise forms of $\tilde{%
\varepsilon}_{i}$ and $\phi_i$ do not affect the qualitative behavior.

The analytical description of the dynamics is most straightforward when the
wavefunction is expanded as \cite{hara,i2019}%
\begin{equation*}
|\psi (t)\rangle =\sum_{n}b_{n}(t)e^{-it\lambda _{n}}\exp (i\langle \varphi
_{n}|\hat{f}|\varphi _{n}\rangle )|\varphi _{n}\rangle \,.
\end{equation*}%
$\lambda _{n}$ and $|\varphi _{n}\rangle $ are the eigenfrequencies and
eigenstates (dressed-states) of the unperturbed Hamiltonian $\hat{H}%
_{0}\equiv \hat{H}[\varepsilon =\chi =\tilde{\varepsilon}_{i}=0]$, where the
index $n$ increases with energy. $b_{n}$ is the slowly varying probability
amplitude and
\begin{eqnarray*}
\hat{f} &\equiv &\varepsilon \hat{n}\frac{\cos \eta t-1}{\eta }-(\hat{a}%
^{\dagger 2}-\hat{a}^{2})\frac{i\varepsilon \eta \sin \eta t}{4\eta \nu } \\
&&+\sum_{k,l>k}\tilde{\varepsilon}_{k,l}\hat{D}_{k,l}\frac{\cos (\eta t+\phi
_{k,l})-\cos \phi _{k,l}}{\eta }
\end{eqnarray*}%
is an operator that will not influence the final results under the carried
assumptions. In the low-excitation regime, $\varepsilon \langle \hat{n}%
\rangle \ll \nu $, the time evolution is given by%
\begin{equation}
\dot{b}_{m}\approx \sum_{n<m}\Theta _{n;m}^{\ast }e^{it(\Delta _{nm}-\eta
)}b_{n}-\sum_{n>m}\Theta _{m;n}e^{-it(\Delta _{nm}-\eta )}b_{n},
\end{equation}%
where $\Delta _{nm}\equiv |\lambda _{n}-\lambda _{m}|$ is the transition
frequency between the states $\varphi _{n}$ and $\varphi _{m}$ and the
corresponding transition rate is
\begin{equation}
\Theta _{m;n}\equiv \frac{\varepsilon }{2}\left[ C_{m;n}+\sum_{k,l>k}\frac{%
\tilde{\varepsilon}_{k,l}\exp (i\phi _{k,l})}{\varepsilon }A_{m;n}^{k,l}%
\right] \,.  \label{q}
\end{equation}%
Thus, the general problem has been reduced to evaluation of the matrix
elements%
\begin{eqnarray}
C_{m;n}&\equiv& \langle \varphi _{m}|\left[ \hat{n}+({\eta }/{4\nu })(\hat{a}%
^{2}-\hat{a}^{\dagger 2})\right] |\varphi _{n}\rangle \notag \\
A_{m;n}^{k,l}&\equiv&
\langle \varphi _{m}|\hat{D}_{k,l}|\varphi _{n}\rangle ~,  \label{xxc}
\end{eqnarray}%
where $C_{m;n}$ ($A_{m;n}^{k,l}$) is the cavity's (atom's)
contribution. It is worth noting that under above approximations a different
functional dependence of $\chi $ would merely modify the prefactor of the
second term in $C_{m;n}$; likewise, a modulation of atomic energies \cite%
{hara,i2019} would be described by an additional matrix element in Eq. (\ref%
{q}).

\section{4-photon DCE with a qubit}

\label{s3}

This section focuses on the case of a qubit with the levels $\{|\mathbf{0}%
\rangle ,|\mathbf{1}\rangle \}$. During DCE the atom should remain in the
same state (the ground state, due to unavoidable relaxation processes), so
it is necessary to operate in the \emph{strong dispersive regime}: $|\nu
-E_{1}|\gg g_{1}\sqrt{m}$ for all the populated cavity Fock states $%
|m\rangle $. Treating the term $g_{1}\hat{D}_{0,1}$ in $\hat{H}_{0}$ via
perturbation theory, one obtains the following (non-normalized) eigenstates
with the atom mainly in the ground state%
\begin{widetext}
\begin{eqnarray*}
|\varphi _{0,k}\rangle &\approx &|\mathbf{0},k\rangle +g_{1}|\mathbf{1}%
\rangle \left[ \frac{\sqrt{k}}{\nu -E_{1}}|k-1\rangle -\frac{\sqrt{k+1}}{\nu
+E_{1}}|k+1\rangle \right] +\frac{g_{1}^{2}}{2\nu }|\mathbf{0}\rangle \left[
\frac{\sqrt{k!/(k-2)!}}{\nu -E_{1}}|k-2\rangle +\frac{\sqrt{(k+2)!/k!}}{\nu
+E_{1}}|k+2\rangle \right] \\
&&+\frac{g_{1}^{3}}{2\nu }|\mathbf{1}\rangle \left[ \frac{\sqrt{k!/(k-3)!}}{%
(3\nu -E_{1})(\nu -E_{1})}|k-3\rangle -\frac{\sqrt{(k+3)!/k!}}{(3\nu
+E_{1})(\nu +E_{1})}|k+3\rangle \right] \\
&+&\frac{g_{1}^{4}}{8\nu ^{2}}|\mathbf{0}\rangle \left[ \frac{\sqrt{k!/(k-4)!%
}}{(3\nu -E_{1})(\nu -E_{1})}|k-4\rangle +\frac{\sqrt{(k+4)!/k!}}{(3\nu
+E_{1})(\nu +E_{1})}|k+4\rangle \right] \,,
\end{eqnarray*}%
where $k\geq 0$. The corresponding eigenfrequencies read%
\begin{eqnarray*}
\lambda _{0,k} &\approx &k\nu +g_{1}^{2}\left[ \frac{k}{\nu -E_{1}}-\frac{k+1%
}{\nu +E_{1}}\right] \\
&&+g_{1}^{4}\left[ \frac{k}{(\nu -E_{1})^{2}}\left( \frac{k-1}{2\nu }-\frac{k%
}{\nu -E_{1}}+\frac{k+1}{\nu +E_{1}}\right) -\frac{k+1}{(\nu +E_{1})^{2}}%
\left( \frac{k+2}{2\nu }+\frac{k}{\nu -E_{1}}-\frac{k+1}{\nu +E_{1}}\right) %
\right].
\end{eqnarray*}%
For the modulation frequency $\eta \approx 4\nu $ the cavity's contribution
to the transition rate between the states $|\varphi _{0,k}\rangle $ and $%
|\varphi _{0,k+4}\rangle $ is%
\begin{equation}
C_{0,k;0,k+4}\approx \frac{3g_{1}^{4}E_{1}\sqrt{(k+1)(k+2)(k+3)(k+4)}}{\nu
(\nu -E_{1})(\nu +E_{1})(3\nu -E_{1})(3\nu +E_{1})}\,.  \label{d}
\end{equation}%
In the dispersive regime this term describes the 4DCE whereby photons are
generated in groups of four. At first sight the transition rate is very
small, being proportional to $(g_{1}/\nu )^{4} $. Fortunately, Eq. (\ref{d})
diverges for $E_{1}\approx 3\nu $ (due to a failure of the above
perturbative expansion), giving a hope that near such 3-photon resonance
\cite{law2} the transition rate might have a peak.

To evaluate the matrix elements in the vicinity of $E_{1}\approx 3\nu $ one
reapplies the perturbation theory choosing as perturbation $g_{1}(\hat{a}%
\hat{\sigma}_{0,1}+\hat{a}^{\dagger }\hat{\sigma}_{1,0})$. Now the
eigenstates with the atom predominantly in the ground state read%
\begin{eqnarray*}
|\varphi _{0,k=0}\rangle &\approx &|\mathbf{0},0\rangle +2g_{1}\left[ \frac{%
s_{2}}{ 3\nu +E_{1}-\beta _{2} }|\alpha _{2}^{-}\rangle -\frac{c_{2}}{ 3\nu
+E_{1}+\beta _{2} }|\alpha _{2}^{+}\rangle \right] \\
&&+\frac{8\sqrt{3}\beta _{2}g_{1}^{2}s_{2}c_{2}}{\left( 3\nu +E_{1}\right)
^{2}-\beta _{2}^{2}}\left[ \frac{s_{4}}{ 7\nu +E_{1}-\beta _{4} }|\alpha
_{4}^{-}\rangle -\frac{c_{4}}{ 7\nu +E_{1}+\beta _{4} }|\alpha
_{4}^{+}\rangle \right] ,
\end{eqnarray*}%
\begin{eqnarray*}
|\varphi _{0,k>2}\rangle &\approx &|\alpha _{k}^{-}\rangle +2g_{1}\left[
\sqrt{k+1}c_{k}\left( \frac{s_{k+2}|\alpha _{k+2}^{-}\rangle }{4\nu +\beta
_{k}-\beta _{k+2}}-\frac{c_{k+2}|\alpha _{k+2}^{+}\rangle }{4\nu +\beta
_{k}+\beta _{k+2}}\right) \right. \\
&&\left. -\sqrt{k-1}s_{k}\left( \frac{s_{k-2}|\alpha _{k-2}^{+}\rangle }{%
4\nu -\beta _{k}-\beta _{k-2}}+\frac{c_{k-2}|\alpha _{k-2}^{-}\rangle }{4\nu
-\beta _{k}+\beta _{k-2}}\right) \right] \\
&&+g_{1}^{2}\left[ \sqrt{\left( k+3\right) \left( k+1\right) }\frac{8\beta
_{k+2}c_{k}c_{k+2}s_{k+2}}{\left( 4\nu +\beta _{k}\right) ^{2}-\beta
_{k+2}^{2}}\left( \frac{s_{k+4}|\alpha _{k+4}^{-}\rangle }{8\nu +\beta
_{k}-\beta _{k+4}}-\frac{c_{k+4}|\alpha _{k+4}^{+}\rangle }{8\nu +\beta
_{k}+\beta _{k+4}}\right) \right. \\
&&+\left( \frac{\left( k+1\right) s_{k+2}^{2}}{4\nu +\beta _{k}-\beta _{k+2}}%
+\frac{\left( k-1\right) s_{k-2}^{2}}{4\nu -\beta _{k}-\beta _{k-2}}+\frac{%
\left( k-1\right) c_{k-2}^{2}}{4\nu -\beta _{k}+\beta _{k-2}}+\frac{\left(
k+1\right) c_{k+2}^{2}}{4\nu +\beta _{k}+\beta _{k+2}}\right) \frac{%
2c_{k}s_{k}|\alpha _{k}^{+}\rangle }{\beta _{k}} \\
&&-(1-\delta _{k,4})\sqrt{\left( k-3\right) \left( k-1\right) }\frac{8\beta
_{k-2}s_{k-2}c_{k-2}s_{k}}{\left( 4\nu -\beta _{k}\right) ^{2}-\beta
_{k-2}^{2}}\left( \frac{s_{k-4}|\alpha _{k-4}^{+}\rangle }{8\nu -\beta
_{k}-\beta _{k-4}}+\frac{c_{k-4}|\alpha _{k-4}^{-}\rangle }{8\nu -\beta
_{k}+\beta _{k-4}}\right) \\
&&\left. -\delta _{k,4}\frac{8\beta _{2}s_{2}c_{2}s_{4}}{7\nu +E_{1}-\beta
_{4}}\frac{g_{1}^{2}\sqrt{3}}{\left( 4\nu -\beta _{4}\right) ^{2}-\beta
_{2}^{2}}|\mathbf{0},0\rangle \right] \,.
\end{eqnarray*}%
Here $s_{k}=\sin \theta _{k}$, $c_{k}=\cos \theta _{k}$, $\theta
_{k}=\arctan [(\nu -E_{1}+\beta _{k})/2g_{1}\sqrt{k}]$, $\beta _{k}=[(\nu
-E_{1})^{2}+4g_{1}^{2}k]^{1/2}$ \cite{diego} and%
\begin{equation*}
|\alpha _{k}^{+}\rangle \equiv s_{k}|\mathbf{0},k\rangle +c_{k}|\mathbf{1}%
,k-1\rangle ~,~|\alpha _{k}^{-}\rangle \equiv c_{k}|\mathbf{0},k\rangle
-s_{k}|\mathbf{1},k-1\rangle \,.
\end{equation*}%
For $\eta \approx 4\nu $ the relevant matrix elements become%
\begin{equation}
C_{0,k;0,k+4}\approx \frac{g_{1}^{4}}{4\nu \left( \nu -E_{1}\right) ^{2}}%
\sqrt{\frac{\left( k+4\right) !}{k!}}\left[ 2\frac{1+8\nu /\left( \nu
-E_{1}\right) }{4\nu -\beta _{k+4}-\beta _{k+2}}-\frac{3}{\nu -E_{1}}-\frac{2%
}{3\nu }-\frac{k}{4\nu }\right]  \label{w}
\end{equation}%
\begin{equation}
A_{0,k;0,k+4}^{0,1}\approx \frac{g_{1}^{3}}{4\nu \left( \nu -E_{1}\right)
^{2}}\sqrt{\frac{\left( k+4\right) !}{k!}}\left[ 1-\frac{8\nu }{4\nu -\beta
_{k+4}-\beta _{k+2}}\right]\,.  \label{c}
\end{equation}%
So the total transition rate, Eq. (\ref{q}), becomes (for the standard dipole
qubit-field interaction, as explained in Sec. \ref{s2})%
\begin{equation}
\Theta _{0,k;0,k+4}\approx \frac{\varepsilon g_{1}^{4}}{8\nu \left( \nu
-E_{1}\right) ^{2}}\sqrt{\frac{\left( k+4\right) !}{k!}}\left[ 2\frac{8\nu
/\left( \nu -E_{1}\right) -1}{4\nu -\beta _{k+4}-\beta _{k+2}}-\frac{3}{\nu
-E_{1}}-\frac{1}{6\nu }-\frac{k}{4\nu }\right] .  \label{t}
\end{equation}

As will be shown in Sec. \ref{s5a}, these expressions are sufficient to
estimate the 4DCE rate in the optimum regime of parameters, despite of a
singularity at $4\nu =\beta _{k+4}+\beta _{k+2}$. This divergence occurs due
to the degeneracy of the states \{$|\alpha _{k+4}^{-}\rangle $,$|\alpha
_{k+2}^{+}\rangle $\}, and can be removed by using the degenerate
perturbation theory with unperturbed states $|\alpha _{k+4}^{-}\rangle \pm
|\alpha _{k+2}^{+}\rangle $, which correspond approximately to $|\mathbf{0}%
,k+4\rangle \pm |\mathbf{1},k+1\rangle$. At the degeneracy point there are
two (non-normalized) eigenstates with the dominant contribution of the state
$|\mathbf{0},k\rangle $, denoted as%
\begin{eqnarray*}
|\varphi _{0,k}^{\pm }\rangle &\approx&|\alpha _{k}^{-}\rangle \pm |\alpha
_{k-2}^{+}\rangle +2g_{1}\left[ \sqrt{k+1}c_{k}\left( \frac{s_{k+2}|\alpha
_{k+2}^{-}\rangle }{4\nu +\beta _{k}-\beta _{k+2}} -\frac{c_{k+2}|\alpha
_{k+2}^{+}\rangle }{4\nu +\beta _{k}+\beta _{k+2}}\right) \right. \\
&&\mp \sqrt{k-1}\left( \frac{s_{k-2}c_{k}|\alpha _{k}^{+}\rangle }{2\beta
_{k}}\pm \frac{c_{k-2}s_{k}|\alpha _{k-2}^{-}\rangle }{4\nu -\beta
_{k}+\beta _{k-2}}\right) \pm \delta _{k,4}\frac{c_{2}|\mathbf{0},0\rangle }{
7\nu +E_{1}-\beta _{4} } \\
&&\left. \pm (1-\delta _{k,4})\sqrt{k-3}c_{k-2}\left( \frac{s_{k-4}|\alpha
_{k-4}^{+}\rangle }{ 8\nu -\beta _{k}-\beta _{k-4} }+\frac{c_{k-4}|\alpha
_{k-4}^{-}\rangle }{ 8\nu -\beta _{k}+\beta _{k-4} }\right) \right] \,.
\end{eqnarray*}%
If for a given value of $E_1$ such degeneracy occurs for the state
containing $|\mathbf{0},k+4\rangle $, then the state containing $|\mathbf{0}%
,k\rangle $ will certainly be nondegenerate (since $\beta _{k+4}+\beta
_{k+2}\ne \beta _{k}+\beta _{k-2}$). Therefore the relevant matrix elements
become
\begin{equation}
\left\vert \langle \varphi _{0,k}|\left( \hat{n}+\hat{a}^{2}-\hat{a}%
^{\dagger 2}\right) |\varphi _{0,k+4}^{\pm }\rangle \right\vert \approx
\frac{3g_{1}\sqrt{k+1}}{4\sqrt{2}\nu }\quad ,\quad \left\vert \langle
\varphi _{0,k}|\hat{D}_{0,1}|\varphi _{0,k+4}^{\pm }\rangle \right\vert =%
\frac{\sqrt{k+1}}{\sqrt{2}}  \label{g2}
\end{equation}%
and the upper bound for Eq. (\ref{t}) is established as $\vert \Theta
_{0,k;0,k+4}^{(\max )}\vert \approx 5\varepsilon g_{1}\sqrt{k+1}/(8\sqrt{2}%
\nu )$.

Therefore the effective 4-photon transition between approximate states $|%
\mathbf{0},k\rangle $ and $|\mathbf{0},k+4\rangle $ can be optimized by
operating in the regime when $\beta _{k+4}+\beta _{k+2}$ is sufficiently
close but not exactly equal to $4\nu $. The corresponding modulation
frequency must be $\eta _{k}\equiv\lambda _{0,k+4}-\lambda _{0,k}$, where
the eigenfrequencies read approximately%
\begin{equation*}
\lambda _{0,k=0}\approx -2g_{1}^{2}\left[ \frac{c_{2}^{2}}{ 3\nu
+E_{1}+\beta _{2} }+\frac{s_{2}^{2}}{ 3\nu +E_{1}-\beta _{2} }\right]
\end{equation*}%
\begin{eqnarray*}
\lambda _{0,k>2} &\approx &\nu \left( k-1/2\right) +\frac{E_{1}}{2}-\frac{1}{%
2}\beta _{k}+2g_{1}^{2}\left[ s_{k}^{2}\left( k-1\right) \left( \frac{%
c_{k-2}^{2}}{4\nu -\beta _{k}+\beta _{k-2}}+\frac{s_{k-2}^{2}}{4\nu -\beta
_{k}-\beta _{k-2}}\right) \right. \\
&&\left. -c_{k}^{2}\left( k+1\right) \left( \frac{c_{k+2}^{2}}{4\nu +\beta
_{k}+\beta _{k+2}}+\frac{s_{k+2}^{2}}{4\nu +\beta _{k}-\beta _{k+2}}\right) %
\right] \,.
\end{eqnarray*}

In the vicinity of $E_{1}\approx 3\nu $ one obtains%
\begin{equation*}
\lambda _{0,k}\approx -\frac{g_{1}^{2}}{4\nu }+\nu \left[ 1-\frac{3g_{1}^{2}%
}{4\nu ^{2}}\right] k+\frac{1}{8}\frac{g_{1}^{4}}{\nu ^{3}}k^{2}
\end{equation*}%
and the resonant modulation frequencies become
\begin{equation}
\eta _{k}\approx 4\nu \left( 1-\frac{3g_{1}^{2}}{4\nu ^{2}}+\frac{g_{1}^{4}}{%
2\nu ^{4}}\right) +\frac{g_{1}^{4}}{\nu ^{3}}k  \label{i}
\end{equation}

To create four photons from the initial zero-excitation state $|\mathbf{0}%
,0\rangle $ the modulation frequency $\eta$ must satisfy the condition $%
|\eta-\eta _{0}|\ll|\Theta_{0,0;0,4}|$. In order to simultaneously couple
the states $\{|\varphi_{{0},4}\rangle ,|\varphi_{{0},8}\rangle \}$ it is
also necessary $|\eta-\eta _{4} |\ll|\Theta_{0,4;0,8}|$, which near $4\nu
=\beta _{4}+\beta _{2}$ requires roughly $\varepsilon /\nu \gg 10\left(
g_{1}/\nu \right) ^{3}$. For a fixed value of $\varepsilon $, on one hand
the small ratio $g_{1}/\nu $ is advantageous for multiple 4-photon
transitions, but on the other hand it lowers the transition rate, increasing
the role of dissipation and other spurious effects. Thus it seems that the
best choice is to work with moderate values of $g_{1}/\nu \sim 0.05$. For
instance, if $g_{1}=0.08\nu $ (bordering the ultra-strong coupling regime)
then multiple 4-photon transitions can take place provided $\varepsilon /\nu
\gg 5\times 10^{-3}$. Therefore, for moderate coupling strengths and $%
\varepsilon /\nu \sim 10^{-2}$ one expects that only the states $|\mathbf{0}%
,4\rangle $ and $|\mathbf{0},8\rangle $ will be significantly populated. In
Sec. \ref{s5a} this prediction will be confirmed numerically.

\section{5-photon DCE with a cyclic qutrit}

\label{s4}

Now the dressed-states of the complete bare Hamiltonian $\hat{H}_{0}$ must be determined. For didactic reasons only the cavity's modulation is considered, since the incorporation of other modulation mechanisms does not affect the qualitative behavior. Far from the resonances $E_{1}\approx
l\nu$  or $E_2\approx l\nu$ with an integer $l$ (the exact conditions will be found shortly) the dressed-states relevant for 5DCE are $|\varphi _{0,k}\rangle =(|\mathbf{0},k\rangle
+\cdots )$ (see \cite{i2019} for the initial terms in the perturbative expansion). After long manipulations one finds for the cavity's
matrix element (\ref{xxc})%
\begin{equation}
C_{0,k;0,k+5}\approx \sqrt{(k+1)(k+2)\cdots (k+5)}\frac{g_{1}g_{2}g_{3}}{\nu
^{3}}\sum_{i=1}^{3}\left( \frac{g_{i}}{\nu }\right) ^{2}Y_{i}\,,  \label{p}
\end{equation}%
where $Y_{i}$ are some complicated dimensionless functions of all the system
parameters $\{g_{i},E_{j},\nu \}$. Therefore, for $g_{1}g_{2}g_{3}\neq 0$
and $\eta \approx 5\nu $ the cavity field can be populated from vacuum via
effective 5-photon transitions, but the transition rate $%
\sim \varepsilon (g_{1}/\nu )^{5}$ is prohibitively small. To explore
the possibility of a resonant enhancement of $C_{m;n}$ one starts
diagonalizing the dominant part of the unperturbed Hamiltonian: $\hat{H}_{0}^{(dom)}/\hbar=\nu \hat{n}+\sum_{k=1}^{2}[E_{k}\hat{\sigma}_{k,k}+%
g_{k}(\hat{a}\hat{\sigma} _{k,k-1}+h.c.)]$. Omitting the
normalization constants (irrelevant for the final approximate results), for $%
n\geq 2$ the approximate eigenstates of $\hat{H}_{0}^{(dom)}$ read%
\begin{equation}
|\mu _{0,n}\rangle =|\mathbf{0},n\rangle +g_{1}\sqrt{n}\left( \frac{\tilde{s}%
_{n-1}^{2}}{D_{n-1,-}}+\frac{\tilde{c}_{n-1}^{2}}{D_{n-1,+}}\right) |\mathbf{%
1},n-1\rangle +\frac{g_{1}\sqrt{n}\tilde{s}_{n-1}\tilde{c}_{n-1}\tilde{\beta}%
_{n-1}}{D_{n-1,+}D_{n-1,-}}|\mathbf{2},n-2\rangle  \label{dd1}
\end{equation}%
\begin{eqnarray}
|\mu _{S,n}\rangle &=&\tilde{s}_{n}|\mathbf{1},n\rangle +\tilde{c}_{n}|%
\mathbf{2},n-1\rangle -\frac{g_{1}\tilde{s}_{n}\sqrt{n+1}}{D_{n,-}}|\mathbf{0%
},n+1\rangle  \notag \\
|\mu _{A,n}\rangle &=&\tilde{c}_{n}|\mathbf{1},n\rangle -\tilde{s}_{n}|%
\mathbf{2},n-1\rangle -\frac{g_{1}\tilde{c}_{n}\sqrt{n+1}}{D_{n,+}}|\mathbf{0%
},n+1\rangle ,  \label{dd2}
\end{eqnarray}%
where $\tilde{s}_{k}=\sin \tilde{\theta}_{k}$,~$\tilde{c}_{k}=\cos \tilde{%
\theta}_{k}$, $\tilde{\theta}_{k}=\tan ^{-1}[(\Delta _{2}+\tilde{\beta}%
_{k})/(2g_{2}\sqrt{k})]$, $\tilde{\beta}_{k}=\sqrt{\Delta
_{2}^{2}+4g_{2}^{2}k}$, $D_{k,\pm }=\nu -E_{1}+\Delta _{2}/2\pm \tilde{\beta}%
_{k}/2$ and $\Delta _{2}\equiv\nu -\left( E_{2}-E_{1}\right) $. The corresponding
eigenfrequencies read%
\begin{eqnarray}
\tilde{\lambda}_{0,n} &=&\nu n+g_{1}^{2}n\left( \frac{\tilde{s}_{n-1}^{2}}{%
D_{n-1,-}}+\frac{\tilde{c}_{n-1}^{2}}{D_{n-1,+}}\right) \left[
1-g_{1}^{2}n\left( \frac{\tilde{s}_{n-1}^{2}}{D_{n-1,-}^{2}}+\frac{\tilde{c}%
_{n-1}^{2}}{D_{n-1,+}^{2}}\right) \right]  \label{k1} \\
\tilde{\lambda}_{S,n} &=&\nu n+E_{1}-\frac{\Delta _{2}}{2}+\frac{\tilde{\beta%
}_{n}}{2}-\frac{\tilde{s}_{n}^{2}g_{1}^{2}\left( n+1\right) }{D_{n,-}}%
+\left( \frac{\tilde{c}_{n}^{2}}{\beta _{n}}+\frac{\tilde{s}_{n}^{2}}{D_{n,-}%
}\right) \frac{\tilde{s}_{n}^{2}g_{1}^{4}\left( n+1\right) ^{2}}{D_{n,-}^{2}}
\notag \\
\tilde{\lambda}_{A,n} &=&\nu n+E_{1}-\frac{\Delta _{2}}{2}-\frac{\tilde{\beta%
}_{n}}{2}-\frac{\tilde{c}_{n}^{2}g_{1}^{2}\left( n+1\right) }{D_{n,+}}%
-\left( \frac{\tilde{s}_{n}^{2}}{\beta _{n}}-\frac{\tilde{c}_{n}^{2}}{D_{n,+}%
}\right) \frac{\tilde{c}_{n}^{2}g_{1}^{4}\left( n+1\right) ^{2}}{D_{n,+}^{2}}.
\label{k2}
\end{eqnarray}
\end{widetext}

The true eigenstates of $\hat{H}_{0}$ can now be obtained by applying the
perturbation theory with the perturbation $\hat{V}= \hat{H}_{0}-\hat{H}_{0}^{(dom)}$. The matrix elements $C_{m;n}$ between the dressed-states with the
atom predominantly in the ground state $|\mathbf{0}\rangle $ will exhibit resonant peaks when
the perturbative corrections to the eigenstate (\ref{dd1})
have poles. This takes place when $\tilde{\lambda}_{0,n}=\tilde{\lambda}%
_{S,n-l}$ or $\tilde{\lambda}_{0,n}=\tilde{\lambda}_{A,n-l}$. The case $l=1$
corresponds to the standard one-photon resonance $E_{1}\approx \nu $, which is not suitable for 5DCE due to a significant excitation of the atom. Hence the regions of {\em resonant enhancement} of the transition rate
between the approximate states $|\mathbf{0},0\rangle $ and $|\mathbf{0}%
,n\rangle $ lie in the vicinity of the constraints $\tilde{\lambda}_{0,n}=%
\tilde{\lambda}_{S,n-l}$ or $\tilde{\lambda}_{0,n}=\tilde{\lambda}_{A,n-l}$
with $l\geq 2$. To benefit from such resonant enhancement while maintaining the atom in the ground state it is necessary to operate in the tail regions of the peaks, where the transition rate decreases by roughly one order of magnitude. Hence one can quantify the strength of 5DCE by calculating
the peak values of $C_{0,k;0,k+5}$. At the exact
resonances the (non-normalized) dressed-states of the Hamiltonian $\hat{H}%
_{0}$ become%
\begin{equation}
|\Lambda _{\pm ,n,l}^{\left( S/A\right) }\rangle=\frac{|\mu _{0,n}\rangle \pm
|\mu _{S/A,n-l}\rangle }{\sqrt{2}}+|\delta \Lambda _{\pm ,n,l}^{\left(
S/A\right) }\rangle  \label{k3},
\end{equation}%
where $|\delta \Lambda _{\pm ,n,l}^{\left( S/A\right) }\rangle $ can be
found from the nondegenerate perturbation theory [since $(\langle \mu
_{0,n}|-\langle \mu _{S/A,n-l}|)\hat{V}(|\mu _{0,n}\rangle +|\mu
_{S/A,n-l})=0$]. For the resonance $\tilde{\lambda}_{0,5}=\tilde{\lambda}%
_{S,5-l}$ (neglecting the small corrections in the eigenfrequencies due to $\hat{V}$) the maximum value of the 5-photon matrix element  is%
\begin{equation}
C_{0,0;0,5}^{(\max )}=\frac{5g_{1}\tilde{\beta}_{1}\tilde{c}_{1}\tilde{s}_{1}%
\tilde{c}_{3}}{36\nu ^{2}}~\text{, for }l=2  \label{S2}
\end{equation}%
\begin{equation}
C_{0,0;0,5}^{(\max )}=\frac{g_{3}\tilde{c}_{2}}{\sqrt{2}}\left[ \tilde{s}%
_{2}^{2}\left( \frac{1}{5\nu -\tilde{\beta}_{2}}-\frac{2}{5\nu }\right) -%
\frac{\tilde{c}_{2}^{2}}{5\nu }+\frac{5}{6\nu }\right],\, l=3
\label{S3}
\end{equation}%
\begin{equation}
C_{0,0;0,5}^{(\max )}=\frac{g_{1}\tilde{s}_{1}}{\sqrt{2}}\left[ \frac{5}{%
6\nu }-\frac{\tilde{s}_{1}^{2}}{5\nu }-\frac{\tilde{c}_{1}^{2}}{5\nu -\tilde{%
\beta}_{1}}\right],\, l=4 . \label{S4}
\end{equation}%
For the resonance $\tilde{\lambda}_{0,5}=\tilde{\lambda}_{A,5-l}$ one obtains%
\begin{equation}
C_{0,0;0,5}^{(\max )}=\frac{5g_{1}\tilde{\beta}_{1}\tilde{c}_{1}\tilde{s}_{1}%
\tilde{s}_{3}}{36\nu ^{2}}~\text{, for }l=2  \label{A2}
\end{equation}%
\begin{equation}
C_{0,0;0,5}^{(\max )}=\frac{g_{3}\tilde{s}_{2}}{\sqrt{2}}\left[ \tilde{c}%
_{2}^{2}\left( \frac{1}{5\nu +\tilde{\beta}_{2}}-\frac{2}{5\nu }\right) -%
\frac{\tilde{s}_{2}^{2}}{5\nu }+\frac{5}{6\nu }\right] ,\, l=3
\label{A3}
\end{equation}%
\begin{equation}
C_{0,0;0,5}^{(\max )}=\frac{g_{1}\tilde{c}_{1}}{\sqrt{2}}\left[ \frac{5}{%
6\nu }-\frac{\tilde{c}_{1}^{2}}{5\nu }-\frac{\tilde{s}_{1}^{2}}{5\nu +\tilde{%
\beta}_{1}}\right] ,\, l=4 . \label{A4}
\end{equation}%
It will be shown in the next section that these simple expressions are in excellent
agreement with the exact numeric results. In a similar manner one can obtain the maximum values of other matrix elements, which are omitted here for the sake of space.

Finally, it is worth mentioning that the modulation of the cavity frequency in the
presence of a cyclic qutrit also allows for 3DCE (when $\eta \approx 3\nu $)
via approximate transitions $|\mathbf{0},k\rangle \rightarrow |\mathbf{0}%
,k+3\rangle $ (3DCE was originally predicted for the modulation
of atomic energy levels in a stationary cavity \cite{hara}). For the
photon generation from vacuum, the resonant enhancement of the transition
rate occurs in the vicinity of $\tilde{\lambda}_{0,3}=\tilde{\lambda}_{S/A,1}$, and the corresponding
maximum values of the matrix elements read%
\begin{equation*}
C_{0,0;0,3}^{(\max )}=\frac{g_{1}\tilde{s}_{1}}{\sqrt{2}}\left( \frac{3}{%
2\nu }-\frac{\tilde{s}_{1}^{2}}{3\nu }-\frac{\tilde{c}_{1}^{2}}{3\nu -\tilde{%
\beta}_{1}}\right) \text{ for }\tilde{\lambda}_{0,3}=\tilde{\lambda}_{S,1}
\end{equation*}%
\begin{equation*}
C_{0,0;0,3}^{(\max )}=\frac{g_{1}\tilde{c}_{1}}{\sqrt{2}}\left( \frac{3}{%
2\nu }-\frac{\tilde{c}_{1}^{2}}{3\nu }-\frac{\tilde{s}_{1}^{2}}{3\nu +\tilde{%
\beta}_{1}}\right) \text{ for }\tilde{\lambda}_{0,3}=\tilde{\lambda}_{A,1}~.
\end{equation*}%
It is remarkable that by carefully adjusting the atomic frequencies $%
\{E_{1},E_{2}\}$ in the far-detuned regime ($E_{1}> 2\nu $), the optimum transition
rates of 5DCE can become comparable to the typical 3DCE rates.


\section{Numeric results}

\label{s5}

This section is devoted to the numeric evaluation of the system dynamics
according to the complete Hamiltonian (\ref{h0}) for the initial state $|%
\mathbf{0},0\rangle $. For simplicity it is assumed that the
atomic coupling strengths are time-independent, $\tilde{\varepsilon}_{i}=0$.
This does not lessen the generality of the discussion, since the
formulas (\ref{q}), (\ref{c}) and (\ref{g2}) suggest that additional modulation
mechanisms do merely modify the transition
rate. Hence, the determination of the dynamics under the sole modulation of $%
\omega(t)$ is primordial for further studies considering arbitrary
modulations of $G_{i}(t)$ (the relationship between $G_{i}$ and other
parameters largely depends on the concrete implementation of the atom-cavity
system). In all the subsequent simulations it is assumed $\chi =(4\omega
)^{-1}d\omega /dt$ and $\varepsilon =3\times 10^{-2}\nu $.

\subsection{4-photon DCE}

\label{s5a}

\begin{figure}[tbh]
\centering\includegraphics[width=1.\linewidth]{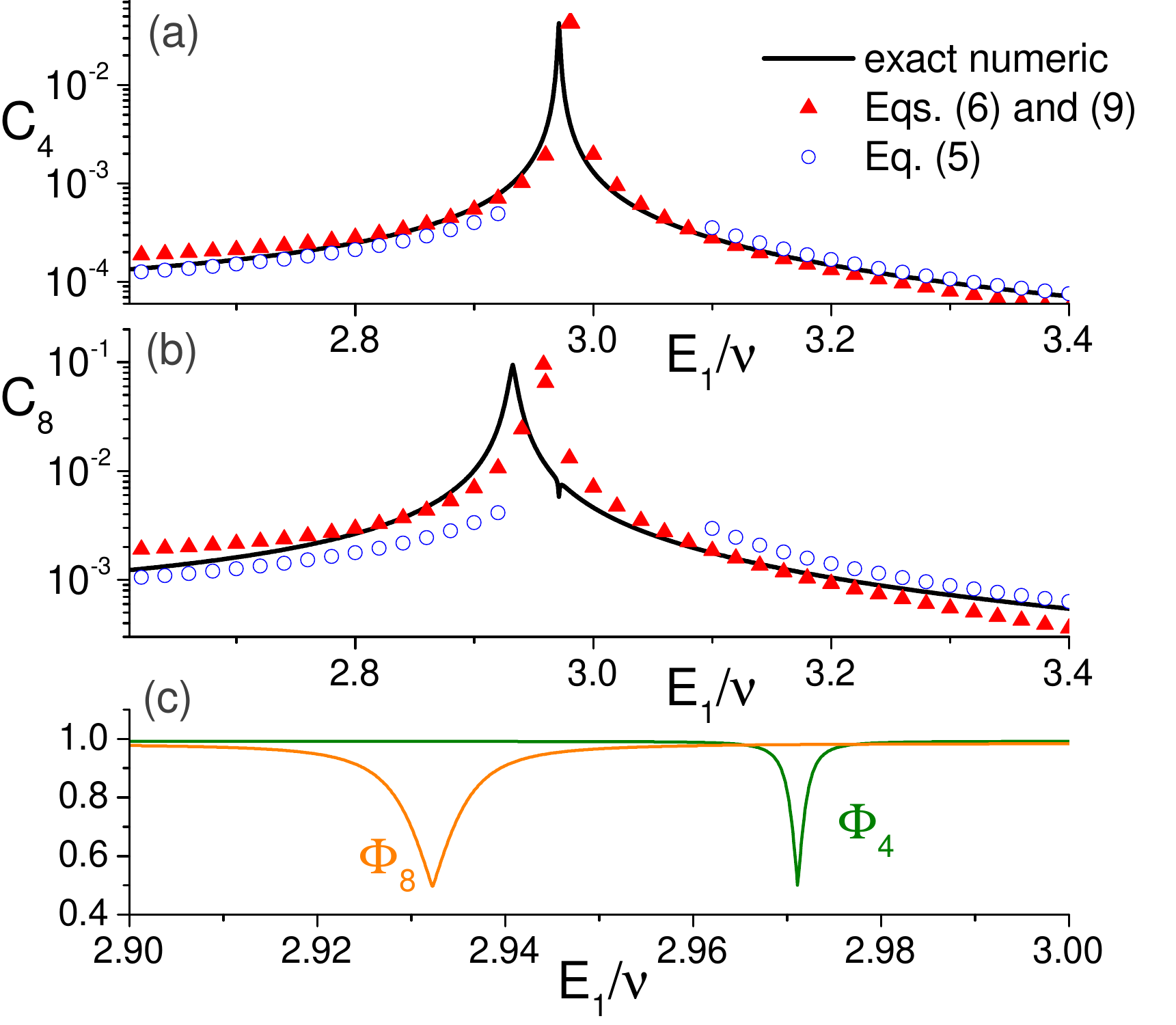}
\caption{(color online) a) Matrix element $C_{4}$ as function of the qubit's
frequency. Solid line represent the exact numeric evaluation; red triangles
correspond to approximate analytic expressions (\protect\ref{w}) and (%
\protect\ref{g2}) valid near the 3-photon resonance $E_{1}\approx 3\protect%
\nu $; blue circles depict Eq. (\protect\ref{d}) valid far from $%
E_{1}\approx 3\protect\nu $. b) Matrix element $C_{8}$. c) Exact numeric
evaluation of the fidelity $\Phi _{m}\equiv |\langle \mathbf{0},m|\protect%
\varphi _{0,m}\rangle |^{2}$ for $m=4$ and $8$.}
\label{f1}
\end{figure}

First it is analyzed the qubit with a realistic \cite{cQED7} coupling
strength $g_{1}=0.08\nu $. The behavior of the two lowest matrix elements $%
C_{4}\equiv |C_{0,0;0,4}|$ and $C_{8}\equiv |C_{0,4;0,8}|$ as function of
the qubit's frequency is shown in Figs. \ref{f1}a-b. The exact values (solid
lines) were obtained through numeric diagonalization of the Hamiltonian $%
\hat{H}_{0}$. Blue circles stand for the analytical formula (\ref{d}) valid
far from the three-photon resonance $E_{1}\approx 3\nu $, and the red
triangles correspond to the expressions (\ref{w}) and (\ref{g2}) applicable
near this resonance. Although the perturbative approach is questionable for
the assumed large ratio $g_{1}/\nu $, there is a good agreement between
exact and approximate results. The main quantitative discrepancy is a slight
displacement in the location of the three-photon resonances, expected
analytically for $4\nu =\beta _{k+4}+\beta _{k+2}$. Fig. \ref{f1}c presents
the exact results for the {\em fidelity} $\Phi _{m}\equiv |\langle \mathbf{0}%
,m|\varphi _{0,m}\rangle |^{2}$ that measures the weight of the state $|%
\mathbf{0},m\rangle $ in the dressed-state $|\varphi _{0,m}\rangle $, for $%
m=4$ and $8$. As expected, in the strong dispersive regime $\Phi _{m}=1/2$
at the three-photon resonances and $\Phi _{m}\approx 1$ otherwise. This confirms that near $E_{1}\approx 3\nu $ it is possible to
implement 4DCE with the vacuum transition rate $|\Theta _{0,0;0,4}|\lesssim
10^{-2}\varepsilon g_{1}/\nu $.

\begin{figure}[tbh]
\centering\includegraphics[width=1.\linewidth]{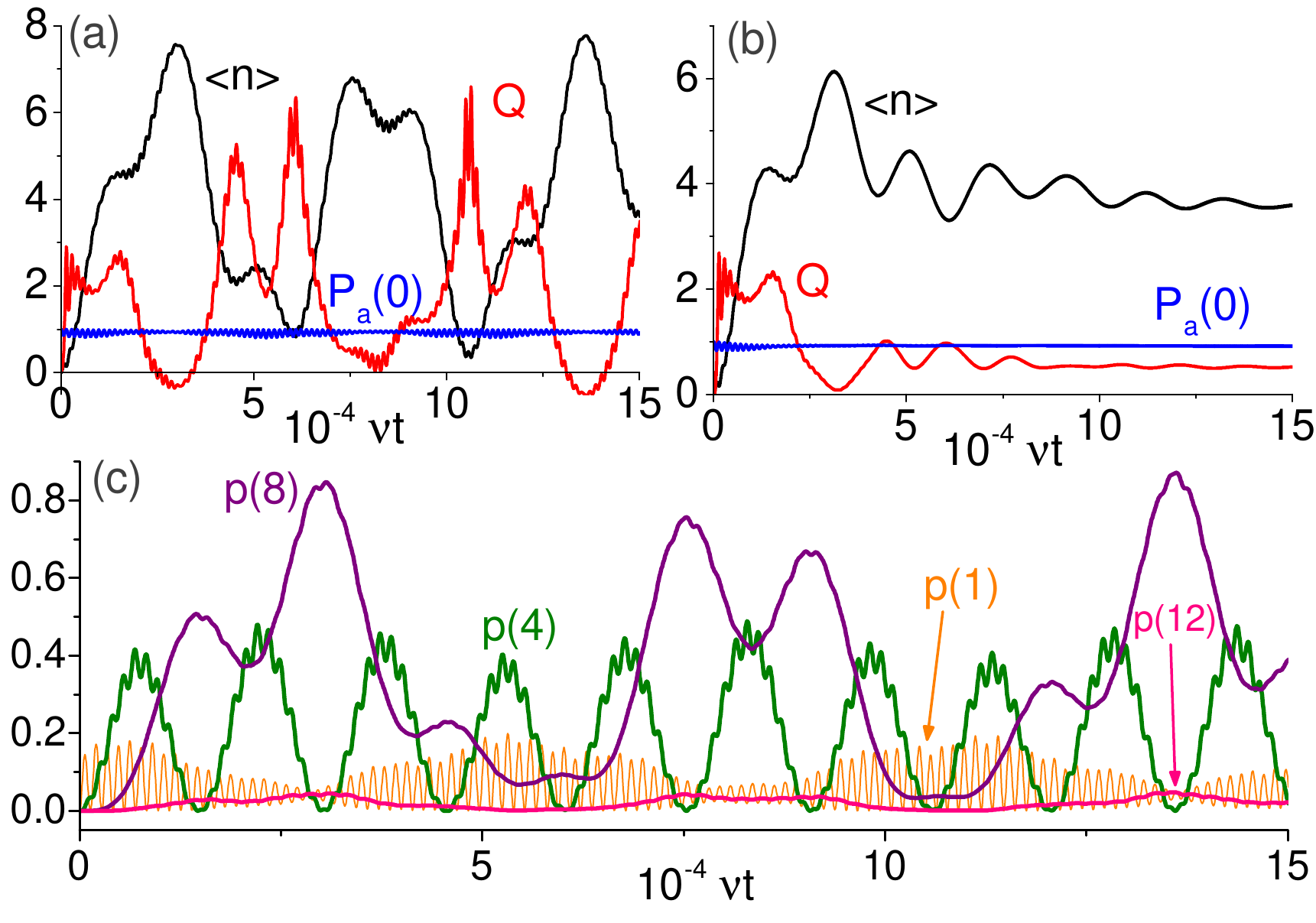}
\caption{(color online) a) Unitary dynamics of 4DCE for parameters $%
E_{1}=2.968\protect\nu $ and $\protect\eta =3.9819\protect\nu $. b) Dynamics
in the presence of weak Markovian dissipation. c) Behavior of the most
populated cavity Fock states under unitary evolution.}
\label{f2}
\end{figure}

Fig. \ref{f2}a illustrates the unitary dynamics for parameters $%
E_{1}=2.968\nu $ and $\eta =3.9819\nu $, obtained by solving numerically the
Schr\"{o}dinger equation. $\langle n\rangle $ is the average photon number, $%
P_{a}(k)$ is the population of the atomic level $|\mathbf{k}\rangle $ and $%
Q=[\langle (\Delta n)^{2}\rangle -\langle n\rangle ]/\langle n\rangle $ is
the Mandel's factor of the cavity field. Several photons are generated from
vacuum via effective four-photon transitions, while the qubit remains mainly
in the ground state. At certain times the $Q$-factor becomes negative,
implying sub-Poissonian field statistics, while at other times it can assume
large ratios, $Q/\langle n\rangle \sim 10$. Such behavior is easily
understood by looking at the evolution of the field in the Fock basis. The
largest photon-number probabilities $p(m)=\mathrm{Tr}[|m\rangle \langle m|%
\hat{\rho}]$ are displayed in Fig. \ref{f2}c, where $\hat{\rho}$ is the
total density operator. $Q<0$ corresponds to the system approximately in the
state $|\mathbf{0},8\rangle $, while the case $Q\gg \langle n\rangle $
occurs when the state $|\mathbf{0},0\rangle $ dominates but there are small
populations of states $|\mathbf{0},8\rangle $ and $|\mathbf{1},1\rangle $.
The denomination \textquotedblleft four-photon dynamical Casimir
effect\textquotedblright\ (4DCE) seems appropriate to describe this
phenomenon, since only the states $|\mathbf{0},4\rangle $, $|\mathbf{0}%
,8\rangle $ and $|\mathbf{0},12\rangle $ become significantly populated,
although the population of the state $|\mathbf{0},12\rangle $ is quite low
due to the effective Kerr nonlinearity [last term in Eq. (\ref{i})]. Due to
the proximity to three-photon resonance, there is also a slight occupation
of the near-degenerate state $|\mathbf{1},1\rangle $, as seen from the
low-amplitude oscillations of $P_{a}(0)$ and $p(1)$.

One can grasp the main qualitative effects of weak dissipation by solving
numerically the phenomenological master equation at zero temperature \cite%
{bla} (see Refs. \cite{diego,palermo} for the discussion on its validity in
similar situations)%
\begin{equation*}
\dot{\rho}=\frac{1}{i\hbar }[\hat{H},\hat{\rho}]+\kappa \mathcal{L}[\hat{a}%
]+\gamma \mathcal{L}[\hat{\sigma}_{0,1}]+\frac{\gamma _{\phi }}{2}\mathcal{L}%
[\hat{\sigma}_{z}]\,\,.
\end{equation*}%
Here $\hat{\sigma}_{z}\equiv \hat{\sigma}_{1,1}-\hat{\sigma}_{0,0}$, $%
\mathcal{L}[\hat{O}]\equiv \hat{O}\hat{\rho}\hat{O}^{\dagger }-\hat{O}%
^{\dagger }\hat{O}\hat{\rho}/2-\hat{\rho}\hat{O}^{\dagger }\hat{O}/2$ is the
Lindblad superoperator, $\kappa $ is the cavity relaxation rate and $\gamma $
($\gamma _{\phi }$) is the atomic relaxation (pure dephasing) rate. Fig. \ref%
{f2}b illustrates the behavior of $\langle n\rangle $, $Q$ and $P_{a}(0)$
for feasible \cite{feas1,feas2} dissipative parameters $\gamma =\gamma
_{\phi }=5\times 10^{-4}g_{1}$ and $\kappa =10^{-4}g_{1}$. The main message
is that several photons can still be generated from vacuum, and for initial
times the dissipative behavior closely resembles the unitary one. For large
times the cavity relaxation leads to excitation of all the Fock states with $%
m\lesssim 10$, so it is not surprising that the behavior is altered
drastically.

\begin{figure}[tbh]
\centering\includegraphics[width=1.\linewidth]{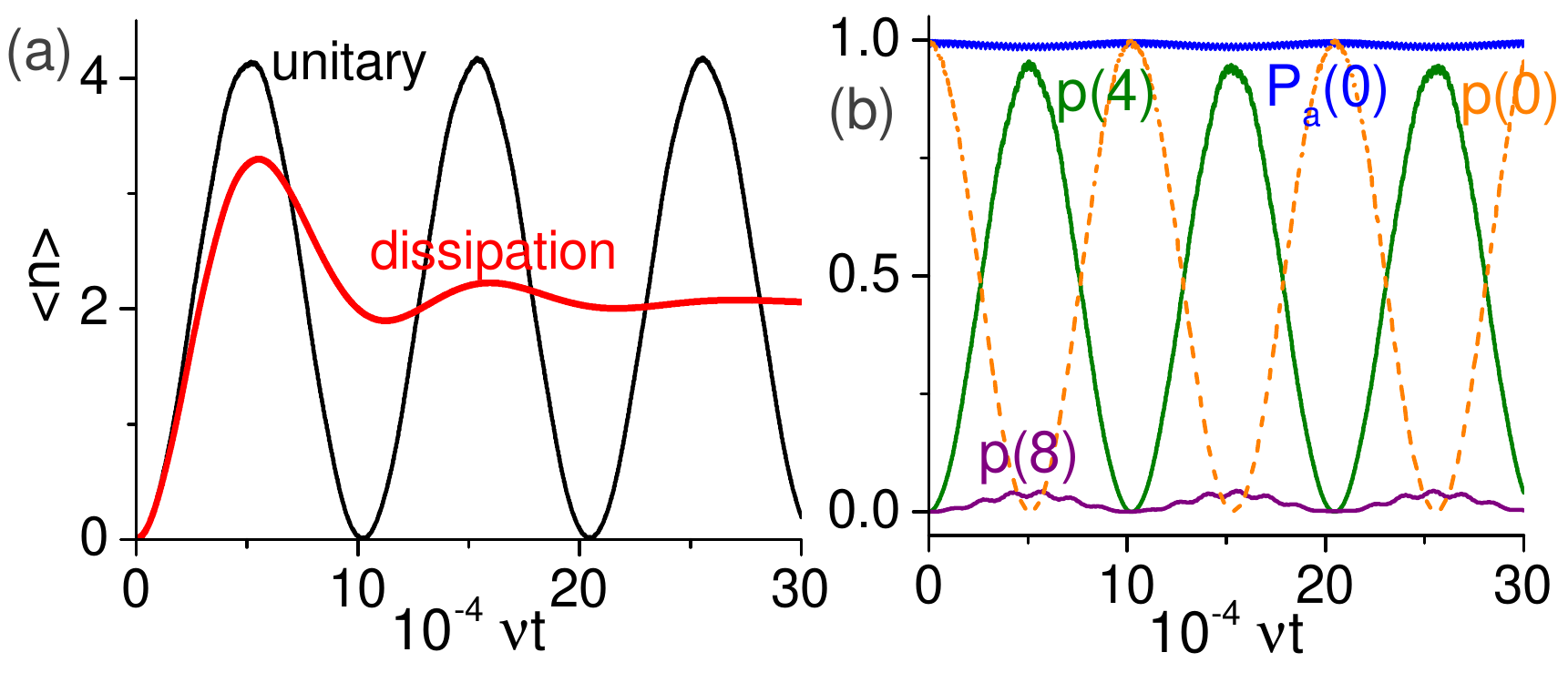}
\caption{(color online) a) Average photon number with and without
dissipation for parameters $E_{1}=2.99\protect\nu $, $\protect\eta %
=3.\,\allowbreak 9821\protect\nu $. b) Ground state population of the qubit,
$P_{a}(0)$, and the largest photon number probabilities $p(k)$ during
unitary evolution.}
\label{f3}
\end{figure}

As was shown in Fig. \ref{f1}, minor changes of $E_{1}$ in the vicinity of
three-photon resonance strongly affect the transition rates. This feature is
illustrated in Fig. \ref{f3}, where the parameters of Fig. \ref{f2} were
slightly changed to $E_{1}=2.99\nu $ and $\eta =3.\,\allowbreak 9821\nu $.
The new behavior is completely different: only the states $|\mathbf{0}%
,0\rangle $ and $|\mathbf{0},4\rangle $ become significantly populated
throughout the evolution; $p(8)$ attains at most 0.04 (making the maximum
value of $\langle n \rangle$ slightly larger than 4), and all other
populations are even smaller. The photon creation is slower than in Fig. \ref%
{f2}, nonetheless, the phenomenon can still occur in the presence of weak
dissipation.

\subsection{5-photon DCE}

\label{s5b}

\begin{figure}[tbh]
\centering\includegraphics[width=1.\linewidth]{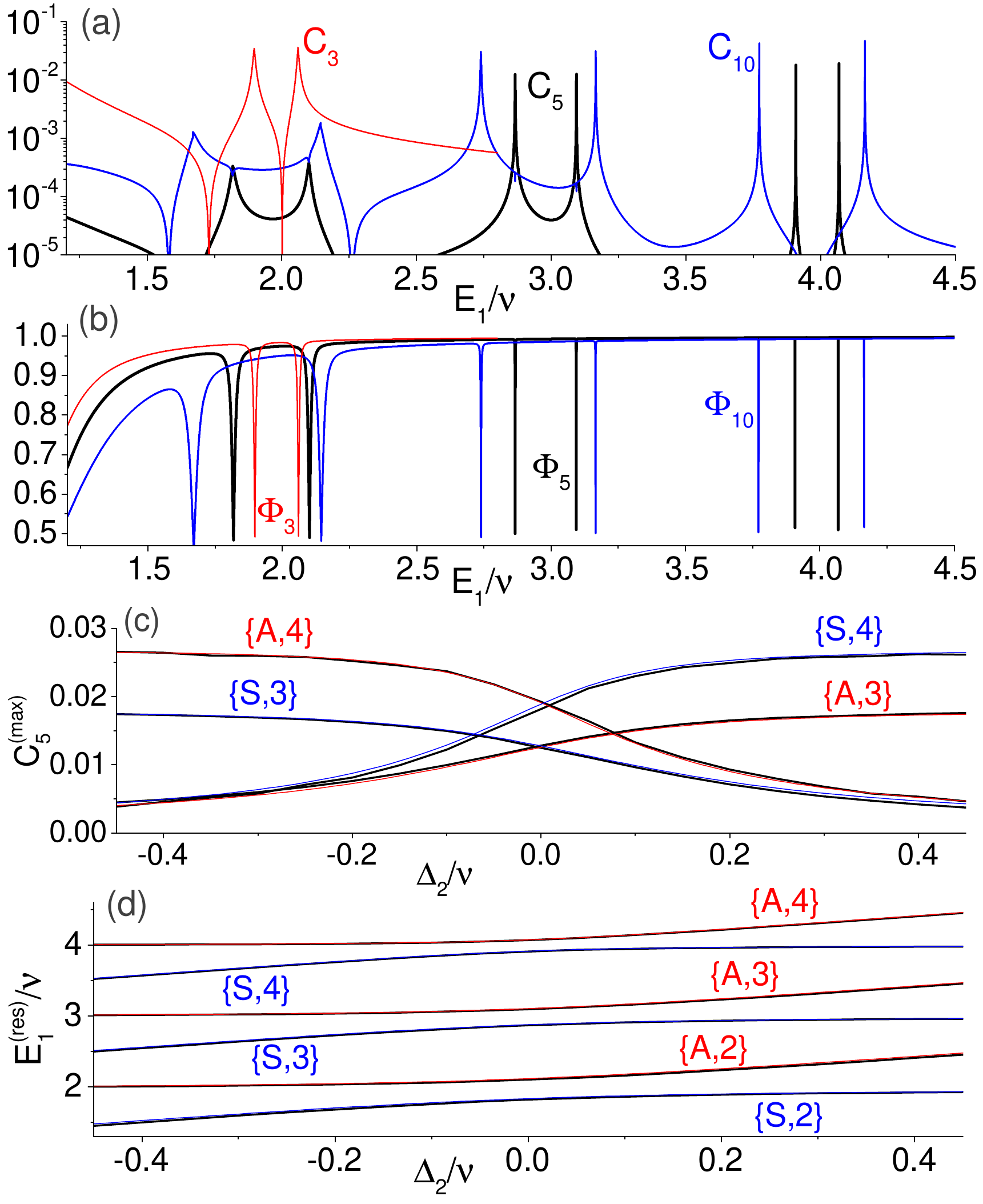}
\caption{(color online) a) Matrix elements $C_{m}$ as function of $E_{1}/%
\protect\nu $ for $\Delta _{2}=0$. b) Fidelities $\Phi _{m}$ for $\Delta
_{2}=0$. c) Peak-values $C_{5}^{(\max )}$ as function of $\Delta _{2}/%
\protect\nu $, where $\{S/A,l\}$ denotes de condition $\tilde{\protect\lambda%
}_{0,5}=\tilde{\protect\lambda}_{S/A,5-l}$. Thick (thin) lines describe the
numeric (analytic) results. d) Values of $E_{1}^{(res)}/\protect\nu $ (for
which $C_{5}$ is maximum) as function of $\Delta _{2}$. }
\label{f4}
\end{figure}

At last the cyclic qutrit is investigated, assuming coupling strengths $%
g_{1}=6\times 10^{-2}\nu $, $g_{2}=8\times 10^{-2}\nu $ and $g_{3}=4\times
10^{-2}\nu $. Fig. \ref{f4}a displays the matrix elements $C_{3}\equiv
|C_{0,0;0,3}|$, $C_{5}\equiv |C_{0,0;0,5}|$ and $C_{10}\equiv |C_{0,5;0,10}|$
as function of $E_{1}/\nu $ for $\Delta _{2}=0$ (for other values of $\Delta
_{2}$ the behavior is qualitatively similar). The dressed-states $|\varphi
_{0,m}\rangle $ (which contain the largest contribution of the state $|%
\mathbf{0},m\rangle $) were obtained via exact numeric diagonalization of
the Hamiltonian $\hat{H}_{0}$. As predicted in Sec. \ref{s4}, the vacuum
transition rate for 5DCE ($C_{5}$) is usually 2 orders of magnitude smaller
than for 3DCE $(C_{3}$). However, near certain atomic frequencies there is a
resonant enhancement of the transition rate, and 5DCE becomes almost as
strong as 3DCE. Fig. \ref{f4}b shows the fidelities $\Phi _{m}$ for $m=3$, $5$ and $10$. As
expected, as long as one stays far from $E_{1}\approx \nu $ and slightly off
the resonance conditions $\tilde{\lambda}_{0,n}=\tilde{\lambda}_{S/A,n-l}$,
the fidelities are very close to 1, allowing for the resonant enhancement of
the transition rate without significantly exciting the atom. It is also
worth noting that the linewidths of the peaks of $C_{m}$ become narrower as $%
E_{1}$ increases, requiring high-precision tuning of the atomic energy
levels in addition to the modulation frequency.

Fig. \ref{f4}c illustrates how the peak-value $C_{5}^{(\max )}$ (associated with 5DCE from vacuum) scales with $%
\Delta _{2}/\nu $, where the atomic energy $E_{1}$ was adjusted according to
the requirement $\tilde{\lambda}_{0,5}=\tilde{\lambda}_{S/A,5-l}$ for $l=3$
and $4$, denoted by the pair of indexes $\{S/A,l\}$ [the case $l=2$ is not
shown since the corresponding rate is one order of magnitude smaller, in
agreement with Eqs. (\ref{S2}) and (\ref{A2})]. Black thick lines denote the
exact numeric results, and the thin lines correspond to Eqs. (\ref{S3}) -- (%
\ref{A4}). The agreement is excellent, and one can see that for the chosen
parameters the optimum transition rates occur for $|\Delta _{2}|\gtrsim
0.2\nu $. Finally, Fig. \ref{f4}d illustrates the dependence of the resonant
value $E_{1}^{(res)}$ that maximizes $C_{5}$ as function of $\Delta _{2}/\nu
$. Analytic results (thin lines) correspond to Eqs. (\ref{k1}) -- (\ref%
{k2}), and are in excellent agreement with the exact numeric results (thick
black lines).

\begin{figure}[tbh]
\centering\includegraphics[width=1.\linewidth]{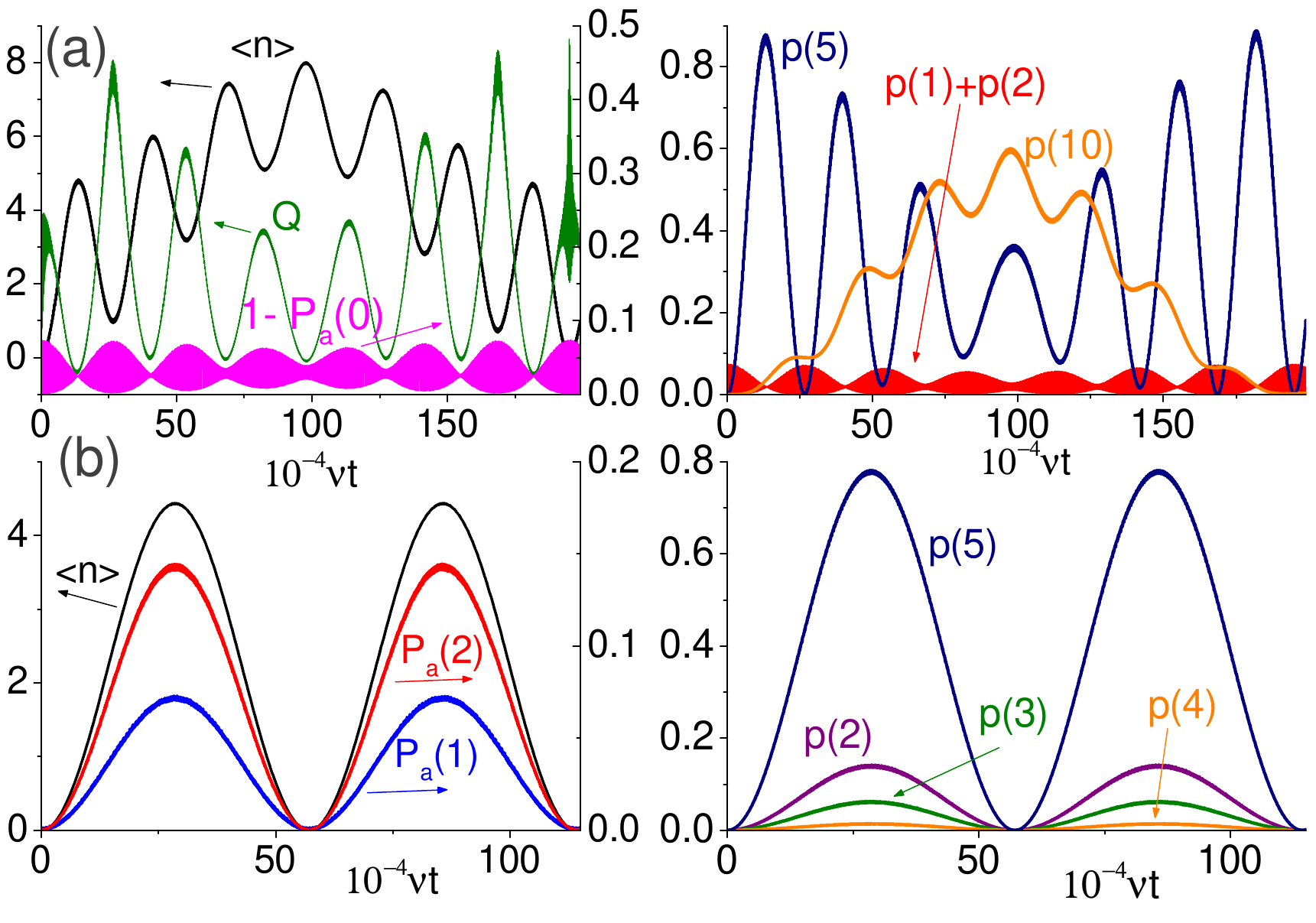}
\caption{(color online) a) 5DCE for parameters $E_{1}=3.105\protect\nu $, $%
E_{2}=4.08\protect\nu $ and $\protect\eta =4.9842\protect\nu $. In the left
panel the scale for $\langle n\rangle $ and \ $Q$ is on the left axis, while
the scale for $1-P_{a}(0)$ is on the right axis. The right panel depicts the
most populated cavity Fock states. b) Similar analysis for $E_{1}=2.2\protect%
\nu $, $E_{2}=3.05\protect\nu $ and $\protect\eta =4.\,\allowbreak 9732%
\protect\nu $, when at most $5$ photons are created.}
\label{f5}
\end{figure}

Actual examples of unitary dynamics are illustrated in Fig. \ref{f5}, as obtained
via numeric solution of the Schr\"{o}dinger equation. In panel \ref{f5}a the
parameters are $E_{1}=3.105\nu $, $E_{2}=4.08\nu $ and $\eta
=4.\,\allowbreak 9842\nu $. The plotted quantities are $\langle n\rangle $, $%
Q$, $1-P_{a}(0)$ and the most populated cavity Fock states. Both $p(1)$ and $%
p(2)$ are very small and have the same order of magnitude, so the quantity $%
p(1)+p(2)$ is plotted instead. In agreement with Eq. (\ref{k3}), $p(1)+p(2)$
almost coincides with $1-P_{a}(0)$, since at the resonance the states $|%
\mathbf{0},5\rangle $, $|\mathbf{1},2\rangle $ and $|\mathbf{2},1\rangle $
are nearly degenerate. The most populated states are $|\mathbf{0},5\rangle $
and $|\mathbf{0},10\rangle $, with other states $|\mathbf{0},5k\rangle $
almost unpopulated due to the mismatch between the modulation frequency $%
\eta $ and $(\lambda _{0,5k}-\lambda _{0,5k-5})$ for $k>2$. In Fig. \ref{f5}%
b similar analysis is carried for $E_{1}=2.2\nu $, $E_{2}=3.05\nu $ and $%
\eta =4.\,\allowbreak 9732\nu $. Now at most $5$ photons are created from
vacuum, but due to the proximity to the resonance peak of $C_5$ the
states $|\mathbf{1},3\rangle $ and $|\mathbf{2},2\rangle $ also become
populated, which explains why $\langle n\rangle $ never attains the value $5$
and why $P_{a}(1)$ and $P_{a}(2)$ deviate significantly from zero.

\section{Conclusions}

\label{s6}

The problem of a single-mode cavity with harmonically modulated frequency
was revisited in the presence of a qubit or a cyclic qutrit. It was found
analytically that the counter-rotating terms in the light-matter interaction
Hamiltonian allow for photon generation from vacuum via effective 4- and
5-photon processes for qubit and cyclic qutrit, respectively, while the
atom remains approximately in the ground state. Usually the associated
transition rates are very small, but they undergo a resonant enhancement by orders of magnitude in the
strong dispersive regime near certain atomic frequencies. For the qubit such
resonance occurs near $E_{1}\approx 3\nu $, while for the qutrit there are six resonance conditions that depend on both $E_1$ and $E_2$. Due to the
effective Kerr nonlinearity, only a limited number of photons can be
generated for a constant modulation frequency. Besides, dissipation alters
drastically the dynamics after some time due to the population of
cavity Fock states forbidden by unitary evolution. Nonetheless, for weak dissipation and sufficiently
strong modulation amplitude, $\varepsilon /\nu \gtrsim 10^{-3}$, 4- and
5-photon DCE could be implemented experimentally for modulation frequencies
$\eta \approx 4\nu $ and $\eta \approx 5\nu $, respectively.

\begin{acknowledgments}
Partial support from Conselho Nacional de Desenvolvimento Cient\'{\i}fico e
Tecnol\'{o}gico -- CNPq (Brazil) is acknowledged.
\end{acknowledgments}


\begin{thebibliography}{99}
\bibitem{asu} The distinction between the instantaneous and fixed
annihilation and creation operators is neglected here, assuming that the
measurements are made after the perturbation has ceased.
\end{thebibliography}
\end{document}